\documentclass[conference]{IEEEtran}
\IEEEoverridecommandlockouts
\usepackage{cite}
\usepackage{amsmath,amssymb,amsfonts}
\usepackage{comment}
\usepackage{algorithmic}
\usepackage{graphicx}
\usepackage{textcomp}
\usepackage{xcolor}
\usepackage{pifont}
\newcommand{\cmark}{\ding{51}}%
\newcommand{\xmark}{\ding{55}}%

\def\BibTeX{{\rm B\kern-.05em{\sc i\kern-.025em b}\kern-.08em
    T\kern-.1667em\lower.7ex\hbox{E}\kern-.125emX}}
\begin{document}

\title{Pre-training Music Classification Models via Music Source Separation
\thanks{This research was supported by the Hellenic Foundation
for Research and Innovation (H.F.R.I.) under the ``3rd Call for H.F.R.I. Research Projects to support Post-Doctoral Researchers'' (Project Acronym: i-MRePlay, Project Number: 7773).}}

\author{Christos Garoufis$^{1,2,3}$, Athanasia Zlatintsi$^{1,2,3}$, and Petros Maragos$^{2,3}$ \\  
\textit{$^1$Institute of Language and Speech Proc., Athena Research Center, Athens, Greece} \\
     \textit{$^2$Institute of Robotics, Athena Research Center, Athens, Greece} \\
     \textit{$^3$School of ECE, National Technical University of Athens, Athens, Greece}\\
     \tt \small{\{christos.garoufis,nancy.zlatintsi\}@athenarc.gr,  maragos@cs.ntua.gr}}

\maketitle

\begin{abstract}
In this paper, we study whether music source separation can be used as a pre-training strategy for music representation learning, targeted at music classification tasks. To this end, we first pre-train U-Net networks under various music source separation objectives, such as the isolation of vocal or instrumental sources from a musical piece; afterwards, we attach a classification network to the pre-trained U-Net and jointly finetune the whole network. The features learned by the separation network are also propagated to the tail network through a convolutional feature adaptation module. Experimental results in two widely used and publicly available datasets indicate that pre-training the U-Nets with a music source separation objective can improve performance compared to both training the whole network from scratch and using the tail network as a standalone in two music classification tasks, music auto-tagging and music genre classification. We also show that our proposed framework can be successfully integrated into both convolutional and Transformer-based backends, highlighting its modularity.

\end{abstract}

\begin{IEEEkeywords}
music source separation, transfer learning, music auto-tagging, music genre classification
\end{IEEEkeywords}

\section{Introduction}
\label{sec:intro}

The recent upsurge in the size of available datasets, as well as their utilization through non-fully supervised learning schemes, is causing a paradigm shift from task-specific approaches to task-agnostic models and representations, suitable for tackling a multitude of inter-related tasks. This trend has transferred to music signal processing as well, with the developed approaches roughly divided in two categories: those that apply transfer learning through models trained in larger datasets, in either similar~\cite{choi17} or not directly related~\cite{castellon21} tasks, and those employing a self-supervised pre-training procedure, using either a single~\cite{spijkervet21, xiao22} or multiple~\cite{huang22, manco22, avramidis23} modalities.

A task that is underexplored for music representation learning, or in conjunction with other tasks in music information retrieval, is music source separation. Music source separation has proven beneficial for artist identification~\cite{sharma19,kim21}, and employed within a joint framework with automatic music transcription~\cite{lin21, cheuk23}, instrument activity detection~\cite{hung20}, and key estimation~\cite{cheuk23}. The association between automatically separated audio sources has also been successfully utilized for audio representation learning~\cite{fonseca21waspaa,garoufis23}. 

A family of music source separation architectures \cite{lin21, jansson17, stoller18, kong21, garoufis2021htmd} makes use of a U-Net \cite{ronneberger2015} structure, which involves an encoder that gradually reduces the input resolution to produce low-dimensional features, and a decoder that recovers the isolated sources by iteratively combining these low-dimensional features. This multi-scale structure has motivated the utilization of U-Nets into neural network architectures tackling other tasks, such as music transcription~\cite{pedersoli20} and music representation learning~\cite{vasquez22}, where stacking a U-Net prior to the rest of the network has resulted in performance improvement.
Apart from the encoder-decoder structure of U-Nets, their suitability for music source separation implies their potential utilization as an either pre-trained, or able to be finetuned, feature extractor for other tasks. Indeed, spectral, timbral, or high-level attributes of musical pieces can be better captured using either excerpts of isolated sources~\cite{garoufis23, deberardinis20}, or recombined low resolution features, containing meaningful high-level information pertaining to these sources~\cite{vasquez22}.

\begin{figure*}[t]
        \centering
        \centerline{\includegraphics[width=18cm]{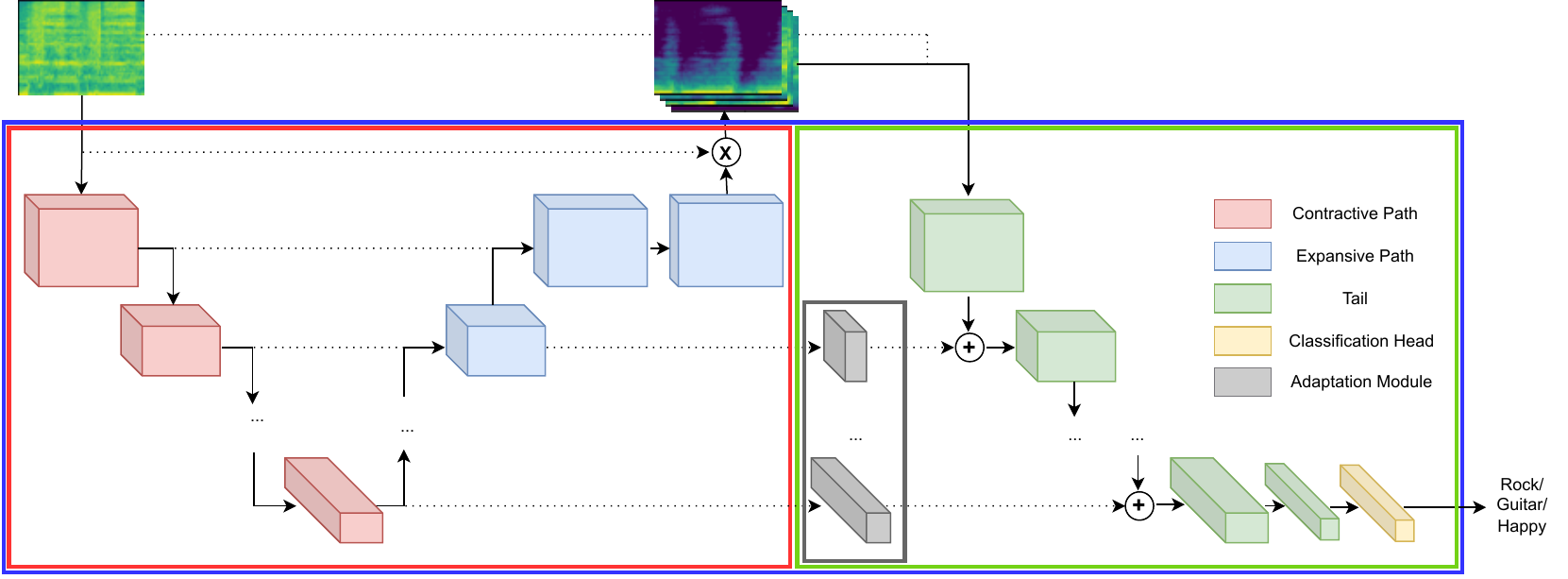}}
    \vspace{-0.4cm}
    \caption{\textcolor{black}{An overview of our proposed framework. We first train a U-Net architecture (red rectangle) with a music source separation objective; then we attach a classification backend (green rectangle) to the pre-trained separation network, using an adaptation module (grey rectangle) for feature adaptation, and train the complete network (blue rectangle) to the desired downstream task, using spectrograms from both the original audio excerpt and the isolated sources.}}
    \vspace{-0.6cm}
\end{figure*}

In this paper\footnote{Code/pretrained models available: https://github.com/cgaroufis/MSSPT}, we investigate whether music source separation can be used for pre-training purposes for other music classification tasks. To this end, driven by the recent success of source separation architectures operating on the Short-Time Fourier Transform (STFT) magnitude~\cite{kong21,luo22}, we pre-train U-Net networks with a source separation objective, and finetune them jointly with a classification backend (\textit{tail}) in downstream tasks, i.e. music auto-tagging and music genre classification. Our approach bears similarities to a recently introduced waveform-domain architecture for self-supervised music representation learning, TUne+ \cite{vasquez22}; however, it operates in a supervised setting in the time-frequency domain, employing a novel pre-training strategy based on music source separation. Experimental results in two datasets, Magna-Tag-A-Tune (MTAT)~\cite{mtat} for music auto-tagging and FMA~\cite{fma16} for music genre classification, with two different classification backends, a VGG-like convolutional network and an Audio Spectrogram Transformer (AST)~\cite{gong21} indicate that training from scratch the classification network with the pre-stacked U-Net does not improve the performance of the classification backends. On the other hand, using a separation objective to pre-train the U-Net can lead to improved results in spite of the smaller size of the pre-training dataset~\cite{musdb18}, indicating the generalizability of our approach to various music classification tasks. We also discuss, for each classification network, the necessity of pre-training it independently to the target downstream task before finetuning it jointly with the U-Net, while showing that the proposed training framework guides the networks towards learning properties of musical signals tied to the musical source used during U-Net pre-training. 

\section{METHODOLOGY}

\subsection{Network Architecture}

The framework utilized, an overview of which is depicted in Fig.~1 for the case of a convolutional backend, is inspired by TUne+, which was recently introduced by Vasquez et al.~\cite{vasquez22} in the context of self-supervised waveform-domain music representation learning. In short, the architecture of TUne+ consists of: a) a \textit{contractive} path (encoder), which gradually downsamples the input through a series of convolutional layers, producing thus multi-resolution features, b) an \textit{expansive} path (decoder), which re-instates the feature map to its initial dimensionality by combining and upsampling the learned contractive features, and c) the \textit{tail}, which produces the final learned embeddings. Since TUne+ was trained within a contrastive learning framework, the output of the tail was used as a representation for downstream tasks. It is also worth noting that feature maps in the expansive path are connected to compatible, dimensionality-wise, feature maps in both the contractive path and the tail through skip connections.

\textbf{Contractive and Expansive Paths}: Both the encoder and the decoder of our network are modeled after the baseline U-Net presented in \cite{kong21}, with the caveat of reducing the filter capacity at half of the original. The encoder receives STFT magnitude spectrograms and consists of 6 blocks, each of which contains 2 2D-convolutional sub-blocks\footnote{Throughout the networks, all convolutional kernels are of dimensionality 3$\times$3, while pooling and upsampling operations are performed at a factor of 2 in each dimension,  unless stated otherwise.}, followed by a max pooling operation. Conversely, the decoder is built symmetrically to the encoder and consists of 6 blocks, comprising transposed convolutional layers followed by convolutional sub-blocks, structured similarly to the encoder, and followed by a final convolutional sub-block. Each decoder block receives the outputs of both the previous decoder block and its symmetric encoder block, propagated through a skip connection.

\textbf{Tail and Classification Head}: Regarding the tail, we experiment with two different network architectures, a) a convolutional network resembling the Short-Chunk CNN audio backend developed in \cite{won20protocol} and b) a Transformer-based backend, based on AST~\cite{gong21}. In both cases, the tail receives both the original input and the output of the U-Net encoder-decoder, and transforms them to the mel scale before further processing, using 128 mel bands. The convolutional backend is built symmetrically to the encoder and the decoder of the network, with 7 2D-convolutional blocks each incorporating 2 convolutional sub-blocks and a max pooling operation; the number of filters in each block is set to the half of~\cite{won20protocol}. On the other hand, the AST backend follows the typical architecture of a Transformer encoder, consisting of 12 Transformer blocks with an embedding dimension of 768, an internal dimension of 3072 and 12 attention heads per block. Before being inserted into the first Transformer block, the input is split into $16 \times 16$ patches, without overlap. Each patch further undergoes a linear projection into an 1D embedding, which is then flattened, summed with a learnable positional embedding and concatenated with a classification token (\texttt{[CLS]}). Following the literature~\cite{gong21,won20protocol}, the classification head is implemented as a linear layer for the AST, receiving the final representation of the classification token, whereas it further includes an intermediate layer of 512 neurons in the case of the CNN.

\textbf{Feature Adaptation Module}: Similar to~\cite{vasquez22}, the features of the expansive path are propagated towards the classification tail. To this end, the dimensions of the feature maps of the expansive path are first aligned with those of the classification tail. After alignment, a 2D-convolution is applied, using an equal number of filters to the channels of the representation in the tail, and the two representations are finally summed. 

In the case of the CNN backend, since the tail is built symmetrically to the expansive path, its feature maps are paired with those of the tail with the same temporal resolution. The spectral dimensions are aligned through a strided 1D max pooling operation, whereas the convolutional layer operates as a linear projection layer, using $1 \times 1$ kernels. On the other hand, the resolution of the embedding sequence remains constant throughout the AST, with the ordering of the embeddings being derived from patchifying the two-dimensional input into an $N_r \times N_c$ grid, which is then flattened. Thus, to produce an embedding sequence with a similar semantically ordering, the 2D convolution is applied with appropriately sized non-overlapping kernels, to generate an $N_r \times N_c$ grid. In this case, the processed feature maps from the expansive path are inserted into the AST at every second layer, after flattening. 

\subsection{Training Procedure}

We devise a three-stage training scheme for training the network. During the first stage, the U-Net composed by the \textit{contractive} and \textit{expansive} paths (red rectangle in Fig.~1) is trained with a source separation objective, where isolated vocal or instrumental sources of the input audio segment are extracted via applying multiplicative soft masks on the input STFT magnitudes. As the loss function, we utilize the time-domain $\mathcal{L}_1$ loss, using the mixture phase, between the ground truth and estimated sources. Then, during the (optional) second stage, the classification network (\textit{tail}) is trained independently in the targeted downstream task until convergence. Finally, the classification network is connected to the trained U-Net via the feature adaptation module, and the complete network (blue rectangle in Fig.~1) is jointly finetuned. During the second and third stages, a supervised loss is applied at the output of the network, whereas the separation loss is discarded.

We note that for architectures involving the AST, we follow the commonly used practice~\cite{gong21,papaioannou23} of initializing the model parameters from an Imagenet-pretrained checkpoint instead of random weight initialization. Similarly, after conducting preliminary experiments, the convolutional filters of the feature adaptation module were initialized using the Imagenet-pretrained weights of the patch embedding layer, after resizing them by bilinear interpolation to fit the respective kernels.

\section{EXPERIMENTAL SETUP}


\textbf{Data \& Preprocessing}: For source separation pre-training, we utilized the musdb18 dataset \cite{musdb18}, a widely used open-access dataset for music source separation. It consists of a total of 150 songs, sampled at 44.1 kHz, with a total duration approximating 10 hours, as well as separate stems for the vocals, drums, bass, and the rest of the instrumental accompaniment (``other''). On the other hand, Magna-Tag-A-Tune (MTAT) \cite{mtat}, as well as the \textit{medium} subset of the Free Music Archive (FMA) dataset~\cite{fma16}, were employed as downstream datasets. MTAT consists of 25,863 29-sec song clips, each associated with a number of tags and sampled at 16 kHz, and has a total duration of approximately 210 hours. Similar to previous work \cite{garoufis23, won20protocol, lee18} we consider the filtered subset of MTAT, which includes only the portion of the initial dataset labelled with at least one of the top-50 tags. The \textit{medium} subset of FMA comprises 25,000 30-sec song segments, at a sampling rate of 44.1 kHz, each grouped into one out of 16 root genres. It reaches a total duration of approximately 208 hours. For all datasets, either the default~\cite{fma16,musdb18} or the commonly-used~\cite{lee18} split between training, validation and testing data was adopted.

As preprocessing, all audio excerpts in musdb18 and FMA were resampled, for compatibility purposes, at 16 kHz. The STFT magnitude of all audio segments was computed using a window length of 512 samples and a hop length of 160, and was log-scaled before further processing.

\textbf{Training and Evaluation Protocol}: Regarding the source separation pre-training, U-Nets were pre-trained for all four uni-source cases, i.e. the extraction of the vocals, bass, drums, or the rest of the accompaniment (``other''), as well as the multi-source case, where source estimates for all four aforementioned sources are provided. The initial learning rate was adjusted separately for each source~\cite{kong21}, whereas additive data augmentation was employed for all sources with the exception of bass. The Adam~\cite{kingma13} optimizer was used with a batch size equal to 8, whereas the learning rate was scheduled after~\cite{kong21}.

During fine-tuning, an appropriately sized segment for each backend (3.85 sec for the CNN, 7.70 sec for the AST) was randomly sampled from each song clip at each epoch and fed to the network. As the optimizer, in the case of the CNN we followed the setup proposed in~\cite{won19}, using a batch size of 16 and applying early stopping via the validation loss, with a patience of 10 epochs. On the other hand, for the architecture incorporating the AST, we use Adam, a batch size of 4 and an initial learning rate of 5e-05, and halve it every 2nd epoch after the 5th one; after 15 epochs, we finetune the networks with two additional epochs using stochastic gradient descent. 

During inference, song-level predictions are obtained by splitting the clips into slices, with overlap equal to half their length, and then averaging the per-slice predictions. For music auto-tagging, the binary cross-entropy was used as the loss function, and the per-tag average ROC-AUC and PR-AUC scores are employed as metrics. On the other hand, for genre classification, the categorical cross-entropy was used as the loss function, and the weighted accuracy (WA) as the metric.

\begin{table*}[t]
    \begin{center}
                \caption{Performance of the trained networks, according to the target source for pre-training the U-Net, in music auto-tagging (MTAT dataset) and music genre classification (FMA dataset). We also compare to discarding the U-Net and using the bare backend (top row), as well as randomly initializing the U-Net (second row). Bold denotes improvement over the tail baseline; 
            $^*$denotes statistically significant improvement ($p < 0.05$) after application of Bonferroni-corrected t-tests.}
            \vspace{-0.15cm}
    \begin{tabular}{|c|c|c||c|c||c||c|c||c|} \cline{4-9}

   \multicolumn{3}{c}{} & \multicolumn {3}{|c||}{CNN Backend} & \multicolumn{3}{c|}{AST Backend} \\ \cline{4-9} \hline U-Net & Pre- & Source(s) & \multicolumn{2}{c||}{MTAT} & FMA & \multicolumn{2}{c||}{MTAT} & FMA  \\ \cline{4-9} & Training & & ROC-AUC & PR-AUC & WA (\%) & ROC-AUC & PR-AUC & WA (\%)  \\ \cline{4-9}
    \hline
        \xmark & - & - & 91.47 $\pm$ 0.03 & 46.50 $\pm$ 0.11 & 66.30 $\pm$ 0.31 & 91.65 $\pm$ 0.05 & 46.82 $\pm$ 0.20 &  67.22 $\pm$ 0.25 \\ \hline
        \cmark & \xmark & -& 91.38 $\pm$ 0.08 & 46.32 $\pm$ 0.35 & \textbf{66.53 $\pm$ 0.13} &  91.58 $\pm$ 0.04 & 46.77 $\pm$ 0.12 & 67.10 $\pm$ 0.61 \\ \hline \hline
        \cmark & \cmark & Bass & 91.45 $\pm$ 0.08 & 46.48 $\pm$ 0.18 & \textbf{66.58 $\pm$ 0.24} & \textbf{91.69 $\pm$ 0.08} & \textbf{47.00 $\pm$ 0.24} & \textbf{67.89 $\pm$ 0.31}\\ \hline
        \cmark & \cmark & Drums & 91.47 $\pm$ 0.07 & \textbf{46.68 $\pm$ 0.20} & \textbf{66.80 $\pm$ 0.19} & 91.57 $\pm$ 0.03 & 46.57 $\pm$ 0.20 & 66.11 $\pm$ 0.46 \\ \hline
        \cmark & \cmark & Other & \textbf{91.59 $\pm$ 0.07} & \textbf{46.98 $\pm$ 0.13}$^*$ & \textbf{67.14 $\pm$ 0.12}$^*$ & \textbf{91.78 $\pm$ 0.12} & \textbf{47.49 $\pm$ 0.23} & 67.09 $\pm$ 0.33 \\ \hline
        \cmark & \cmark & Vocals & \textbf{91.85 $\pm$ 0.03}$^*$ & \textbf{47.16 $\pm$ 0.17}$^*$ & 66.10 $\pm$ 0.41 & \textbf{91.89 $\pm$ 0.08}$^*$ & \textbf{47.21 $\pm$ 0.34} & 66.77 $\pm$ 0.14 \\ \hline \hline
        \cmark & \cmark & Multiple & \textbf{91.50 $\pm$ 0.02 }& \textbf{46.65 $\pm$ 0.12} & \textbf{66.52 $\pm$ 0.21} & \textbf{91.87 $\pm$ 0.08}$^*$ & \textbf{47.31 $\pm$ 0.13}$^*$ & \textbf{67.40 $\pm$ 0.93} \\ \hline 

    \end{tabular}
    \end{center}
    \vspace{-0.75cm}
\end{table*}

\section{RESULTS AND DISCUSSION}

\begin{table}[t]
    \begin{center}
                \caption{Impact of pre-training the classification backend before fine-tuning for the case of accompaniment separation and both CNN and AST-based backends, in music auto-tagging (MTAT).}
                \vspace{-0.15cm}
    \begin{tabular}{|c|c||c|c|}
    
 \hline
   Backend & Pre-Training & ROC-AUC & PR-AUC \\ \hline \hline
   CNN & \xmark & 91.35 $\pm$ 0.15 & 46.15 $\pm$ 0.44  \\ \hline
   CNN & \cmark & \textbf{91.59 $\pm$ 0.07} & \textbf{46.98 $\pm$ 0.13} \\ \hline \hline
  AST & \xmark & \textbf{91.78 $\pm$ 0.12} & \textbf{47.49 $\pm$ 0.23} \\ \hline
  AST & \cmark & 91.47 $\pm$ 0.08 & {46.58 $\pm$ 0.16} \\ \hline
    \end{tabular}
    \end{center}
    \vspace{-0.15cm}
    \vspace{-0.85cm}
\end{table} 

 \textbf{Main Results}: In Table~I, we present the results for each case of source-wise pre-training, for both backends. As baselines, we consider the performance of the \textit{tail} models, without receiving the input, or the features, of the pre-trained separator (first row in Table~I), as well as the complete models, with the \textit{contractive} and \textit{expansive} paths randomly initialized (second row in Table~I). For all cases, we report the mean value and standard error over 5 runs. An initial conclusion is that in contrast to self-supervised learning~\cite{vasquez22}, the inclusion of a randomly initialized U-Net before the network tail does not necessarily guarantee improved performance. On the other hand, pre-training the U-Net in music source separation, despite using a dataset smaller than the downstream ones for pre-training, can prove beneficial for music classification tasks. In particular, most separation objectives lead to improved performance in music auto-tagging, irrespective of the classification backend, with the largest PR-AUC increase achieved for vocal separation in the case of the CNN, and accompaniment separation in the case of AST. Regarding music genre classification, evaluation of the CNN-based architecture leads to better results for the majority of the pre-training objectives; this improvement is statistically significant for accompaniment separation. However, the trend is reversed for the AST backend, where only the bass tracks lead to consistent performance increase as pre-training targets. 

 Regarding the model comparison according to the integrated backend, we observe that the variants combining the pre-trained U-Net with AST slightly outperform, for most source-task configurations, the fully convolutional ones. This is in agreement with both the the recent literature~\cite{papaioannou23} and the performance of the bare backends -- which is close to the metrics reported in~\cite{won20protocol} and ~\cite{papaioannou23}. We also note that, as we will expand on later, the best results (reported in Table I) were obtained through the complete three-stage scheme for the CNN, but without the optional second stage for the AST.

\begin{figure*}[t]
    \vspace{-0.1cm}
        \centering
        \centerline{\includegraphics[width=0.85\linewidth]{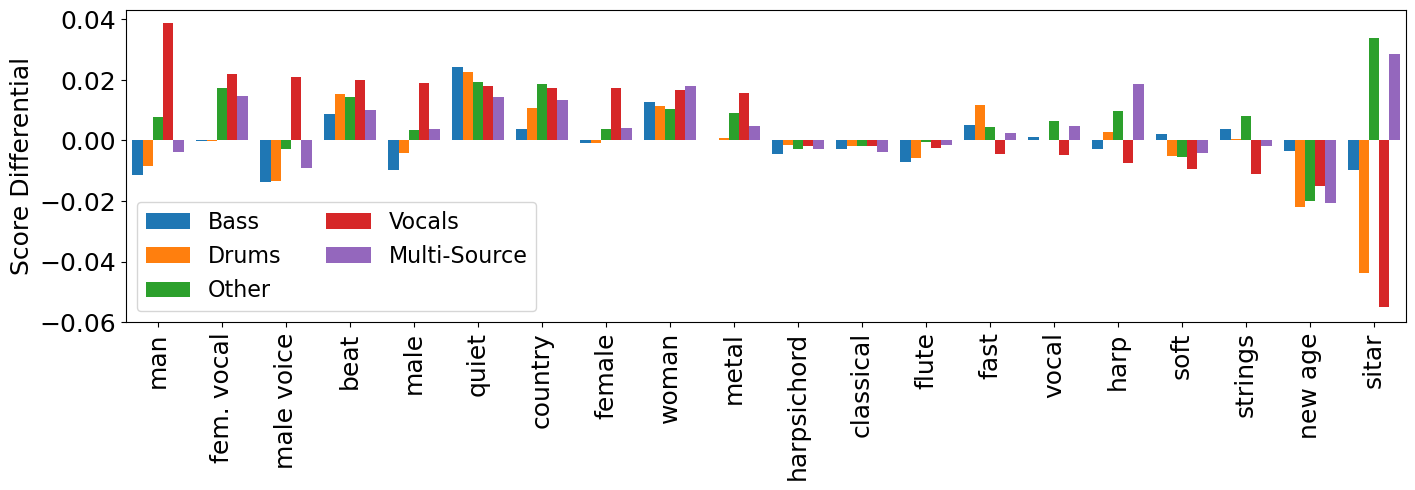}}
    \vspace{-0.3cm}
    \caption{\textcolor{black}{Relative difference (PR-AUC scores, averaged across the CNN-based and AST-based models) between the performance of the end-to-end trained baseline and our proposed framework in selected MTAT tags, according to the objective of the source separation pre-training.}}
    \vspace{-0.6cm}
\end{figure*}
Concerning the performance of individual sources, the results indicate a consistently positive effect of pre-training the U-Net on separating the melodic accompaniment of the musical mixture. On the other hand, while setting the vocals as the target source during pre-training results in worse performance than the tail baselines in music genre classification, it leads to improved performance in the task of music auto-tagging. This result can be correlated to the efficacy of vocal excerpts in identifying artist-specific features, as shown in~\cite{sharma19,kim21}, and as will be further discussed, is linked to the nature of the tags in MTAT. Finally, in contrast to contrastive representation learning~\cite{garoufis23}, using a multi-source pre-training objective does not necessarily outperform single-source pre-training; it is noteworthy, though, that it appears to benefit more from the higher capacity of the AST backend in both downstream tasks.

\textbf{Impact of Backend Pre-Training}: In Table II, we present the ROC-AUC and PR-AUC scores achieved in MTAT for both classification backends, depending on whether the classification backend was independently pre-trained before the joint finetuning phase, for the case of accompaniment separation pre-training. The results indicate that while in the case of the CNN, the network benefits significantly from independent pre-training of the classifcation backend, this technique proves detrimental to the results for the AST backend. 
We hypothesize that the relative lack of inductive biases in Transformer architectures, compared to CNNs, leads to easier network overfitting; additionally, the use of Imagenet-pretrained weights provides a good starting point for the AST, enabling its smooth generalization into other domains.

\textbf{Qualitiative Analysis}: In Fig.~2, we display the tag-wise PR-AUC score difference between the tail baseline and our proposed framework (average between the CNN and AST configurations) for the case of vocal separation pre-training, for the top-10 and bottom-10 MTAT tags in terms of performance differential; for comparative purposes, we also display the performance achieved with the other separation pre-training targets in the same tags. We observe that vocal separation pre-training succeeds in steering the model towards tags related to the vocals, as well as semantic information about them. On the other hand, the tags where the vocally pre-trained model performs worse than the baseline mostly concern recognition of instruments, or genres with sparse vocals.

\section{CONCLUSIONS}

In this work, we investigated whether music source separation can be utilized as a pre-training objective for music classification tasks. Experimental results in two different tasks, music auto-tagging and music genre classification, using both convolutional and Transformer-based backends indicate that our proposed approach can lead to improved performance in both tasks, depending on the target source during pre-training, highlighting its potential applicability in diverse music classification tasks as well as its flexibility. In the future, we would like to exploit more diverse source separation techniques for the pre-training stage~\cite{chen22}, as well as alleviate the computational overhead introduced by network stacking by means of feature-level supervision~\cite{hung22}.
\\
\\
\textbf{Acknowledgments}: We sincerely thank C. Papaioannou for his constructive comments on the content of the paper.

\bibliographystyle{IEEEtran}
\bibliography{eusipco24submission}

\begin{thebibliography}{10}
\providecommand{\url}[1]{#1}
\csname url@samestyle\endcsname
\providecommand{\newblock}{\relax}
\providecommand{\bibinfo}[2]{#2}
\providecommand{\BIBentrySTDinterwordspacing}{\spaceskip=0pt\relax}
\providecommand{\BIBentryALTinterwordstretchfactor}{4}
\providecommand{\BIBentryALTinterwordspacing}{\spaceskip=\fontdimen2\font plus
\BIBentryALTinterwordstretchfactor\fontdimen3\font minus \fontdimen4\font\relax}
\providecommand{\BIBforeignlanguage}[2]{{%
\expandafter\ifx\csname l@#1\endcsname\relax
\typeout{** WARNING: IEEEtran.bst: No hyphenation pattern has been}%
\typeout{** loaded for the language `#1'. Using the pattern for}%
\typeout{** the default language instead.}%
\else
\language=\csname l@#1\endcsname
\fi
#2}}
\providecommand{\BIBdecl}{\relax}
\BIBdecl

\bibitem{choi17}
K.~Choi, G.~Fazekas, M.~Sandler, and K.~Cho, ``{Transfer Learning for Music Classification and Regression Tasks},'' in \emph{Proc. ISMIR 2017}, Suzhou, China, 2017.

\bibitem{castellon21}
R.~Castellon, C.~Donahue, and P.~Liang, ``{Codified Audio Language Modeling Learns Useful Representations for Music Information Retrieval},'' in \emph{Proc. ISMIR 2021}, online, 2021.

\bibitem{spijkervet21}
J.~Spijkervet and J.~A. Burgoyne, ``{Contrastive Learning of Musical Representations},'' in \emph{Proc. ISMIR 2021}, online, 2021.

\bibitem{xiao22}
H.~Zhao, C.~Zhang, B.~Zhu, Z.~Ma, and K.~Zhang, ``{S3T: Self-Supervised Pre-training with Swin Transformer for Music Classification},'' in \emph{Proc. ICASSP 2022}, Singapore, Singapore, 2022.

\bibitem{huang22}
Q.~Huang, A.~Jansen, J.~Lee, R.~Ganti, J.~Y. Li, and D.~P. Ellis, ``{MuLan: A Joint Embedding of Music Audio and Natural Language},'' in \emph{Proc. ISMIR 2022}, Bengaluru, India, 2022.

\bibitem{manco22}
I.~Manco, E.~Benetos, E.~Quinton, and G.~Fazekas, ``{Learning Music Audio Representations via Weak Language Supervision},'' in \emph{Proc. ICASSP 2022}, Singapore, Singapore, 2022.

\bibitem{avramidis23}
K.~Avramidis, S.~Stewart, and S.~Narayanan, ``{On the Role of Visual Context in Enriching Music Representations},'' in \emph{Proc. ICASSP 2023}, Rhodes, Greece, 2023.

\bibitem{sharma19}
B.~Sharma, R.~K. Das, and H.~Li, ``{On the Importance of Audio-Source Separation for Singer Identification in Polyphonic Music},'' in \emph{Proc. Interspeech 2019}, Graz, Austria, 2019.

\bibitem{kim21}
K.~Kim, J.~Lee, S.~Kum, and J.~Nam, ``{Learning a Cross-Domain Embedding Space of Vocal and Mixed Audio with a Structure-Preserving Triplet Loss},'' in \emph{Proc. ISMIR 2021}, online, 2021.

\bibitem{lin21}
L.~Lin, Q.~Kong, J.~Jiang, and G.~Xia, ``{A Unified Model for Zero-Shot Music Source Separation, Transcription and Synthesis},'' in \emph{Proc. ISMIR 2021}, online, 2021.

\bibitem{cheuk23}
K.~W. Cheuk, K.~Choi, Q.~Kong \emph{et~al.}, ``{Jointist: Simultaneous Improvement of Multi-instrument Transcription and Music Source Separation via Joint Training},'' \emph{arXiv preprint arXiv:2302.00286}, 2023.

\bibitem{hung20}
Y.-N. Hung and A.~Lerch, ``{Multitask Learning for Instrument Activation Aware Music Source Separation},'' in \emph{Proc. ISMIR 2020}, online, 2020.

\bibitem{fonseca21waspaa}
E.~Fonseca, A.~Jansen, D.~P. Ellis, S.~Wisdom \emph{et~al.}, ``{Self-Supervised Learning from Automatically Separated Sound Scenes},'' in \emph{Proc. WASPAA 2021}, New Waltz, USA, 2021.

\bibitem{garoufis23}
C.~Garoufis, A.~Zlatintsi, and P.~Maragos, ``{Multi-Source Contrastive Learning from Musical Audio},'' in \emph{Proc. SMC 2023}, Stockholm, Sweden, 2023.

\bibitem{jansson17}
A.~Jansson, E.~Humphrey, N.~Montecchio, R.~Bittner, A.~Kumar, and T.~Weyde, ``{Singing Voice Separation with Deep U-Net Convolutional Networks},'' in \emph{Proc. ISMIR 2017}, Suzhou, China, 2017.

\bibitem{stoller18}
D.~Stoller, S.~Ewert, and S.~Dixon, ``{Wave-U-Net: A Multi-Scale Neural Network for End-to-End Audio Source Separation},'' in \emph{Proc. ISMIR 2018}, Paris, France, 2018.

\bibitem{kong21}
Q.~Kong, Y.~Cao, H.~Liu, K.~Choi, and Y.~Wang, ``{Decoupling Magnitude and Phase Estimation with Deep Res-U-Net for Music Source Separation},'' in \emph{Proc. ISMIR 2021}, online, 2021.

\bibitem{garoufis2021htmd}
C.~Garoufis, A.~Zlatintsi, and P.~Maragos, ``{HTMD-Net: A Hybrid Masking-Denoising Approach to Time-Domain Monaural Singing Voice Separation},'' in \emph{Proc. EUSIPCO 2021}, online, 2021.

\bibitem{ronneberger2015}
O.~Ronneberger, P.~Fischer, and T.~Brox, ``{U-Net: Convolutional Networks for Biomedical Image Segmentation},'' in \emph{Proc. MICCAI 2015}, Munich, Germany, 2015.

\bibitem{pedersoli20}
F.~Pedersoli, G.~Tzanetakis, and K.~M. Yi, ``{Improving Music Transcription by Pre-Stacking a U-Net},'' in \emph{Proc. ICASSP 2020}, online, 2020.

\bibitem{vasquez22}
M.~V{\'a}squez and J.~Burgoyne, ``{Tailed U-Net: Multi-Scale Music Representation Learning},'' in \emph{Proc. ISMIR 2022}, Bengaluru, India, 2022.

\bibitem{deberardinis20}
J.~De~Berardinis, A.~Cangelosi, and E.~Coutinho, ``{The Multiple Voices of Musical Emotions: Source Separation for Improving Music Emotion Recognition Models and their Interpretability},'' in \emph{Proc. ISMIR 2020}, online, 2020.

\bibitem{luo22}
Y.~Luo and J.~Yu, ``{Music Source Separation with Band-Split RNN},'' \emph{IEEE/ACM Transactions on Audio, Speech, and Language Processing}, vol.~31, pp. 1893--1901, 2023.

\bibitem{mtat}
E.~Law, K.~West, M.~I. Mandel, M.~Bay, and J.~S. Downie, ``{Evaluation of Algorithms using Games: The Case of Music Tagging},'' in \emph{Proc. ISMIR 2009}, Kobe, Japan, 2009.

\bibitem{fma16}
M.~Defferrard, K.~Benzi, P.~Vandergheynst, and X.~Bresson, ``{FMA: A Dataset for Music Analysis},'' in \emph{Proc. ISMIR 2017}, Suzhou, China, 2017.

\bibitem{gong21}
Y.~Gong, Y.-A. Chung, and J.~Glass, ``{AST: Audio Spectrogram Transformer},'' in \emph{Proc. Interspeech 2021}, Brno, Czechia, 2021.

\bibitem{musdb18}
Z.~Rafii, A.~Liutkus, F.-R. St{\"o}ter, S.~I. Mimilakis, and R.~Bittner, ``{The MUSDB18 Corpus for Music Separation},'' https://doi.org/10.5281/zenodo.1117372, 2017.

\bibitem{won20protocol}
M.~Won, A.~Ferraro, D.~Bogdanov, and X.~Serra, ``{Evaluation of CNN-Based Automatic Music Tagging Models},'' in \emph{Proc. SMC 2020}, online, 2020.

\bibitem{papaioannou23}
C.~Papaioannou, E.~Benetos, and A.~Potamianos, ``{From West to East: Who Can Understand the Music of the Others Better?}'' in \emph{Proc. ISMIR 2023}, Milan, Italy, 2023.

\bibitem{lee18}
J.~Lee, J.~Park, K.~L. Kim, and J.~Nam, ``{SampleCNN: End-to-end Deep Convolutional Neural Networks using Very Small Filters for Music Classification},'' \emph{Applied Sciences}, vol.~8, no.~1, p. 150, 2018.

\bibitem{kingma13}
D.~Kingma and J.~Ba, ``{Adam: A Method for Stochastic Optimization},'' in \emph{Proc. ICLR 2015}, San Diego, CA, USA, 2015.

\bibitem{won19}
M.~Won, S.~Chun, and X.~Serra, ``{Toward Interpretable Music Tagging with Self-Attention},'' \emph{arXiv preprint arXiv:1906.04972}, 2019.

\bibitem{chen22}
K.~Chen, X.~Du, B.~Zhu, Z.~Ma, T.~Berg-Kirkpatrick, and S.~Dubnov, ``{Zero-shot Audio Source Separation Through Query-Based Learning from Weakly-Labeled Data},'' in \emph{Proc. AAAI 2022}, Vancouver, BC, Canada, 2022.

\bibitem{hung22}
Y.-N. Hung and A.~Lerch, ``{Feature-Informed Embedding Space Regularization for Audio Classification},'' in \emph{Proc. EUSIPCO 2022}, Belgrade, Serbia, 2022.

\end{thebibliography}

\end{document}